\documentclass[12pt]{article}

\usepackage{amssymb}
\usepackage{amsmath}
\usepackage{amscd}
\usepackage{latexsym}
\usepackage{graphicx}

\usepackage{cite}

\newcommand{\be}{\begin{equation}}
\newcommand{\ee}{\end{equation}}

\newcommand{\dlt}{\delta}

\newcommand{\bt}{\beta}

\newcommand{\ep}{\varepsilon}
\newcommand{\al}{\alpha}
\newcommand{\ra}{\rightarrow}

\newcommand{\cL}{{\cal L}}

\newcommand{\cH}{{\cal H}}

\newcommand{\rgl}{\rangle}
\newcommand{\lgl}{\langle}

\begin{document}

\begin{center}

{\Large {\bf Quantum probabilities of composite events in quantum measurements
with multimode states} \\ [5mm]

V.I. Yukalov$^{1,2}$ and D. Sornette$^{1,3}$ } \\ [3mm]

{\it
$^1$ D-MTEC, ETH Z\"urich (Swiss Federal Institute of Technology)\\
Z\"urich CH-8092, Switzerland \\ [2mm]

$^2$Bogolubov Laboratory of Theoretical Physics, \\
Joint Institute for Nuclear Research, Dubna 141980, Russia \\ [2mm]

$^3$ Swiss Finance Institute, \\
c/o University of Geneva, CH-1211 Geneva 4, Switzerland }

\end{center}

\vskip 3cm

\begin{abstract}
The problem of defining quantum probabilities of composite events is
considered. This problem is of high importance for the theory of quantum
measurements and for quantum decision theory that is a part of measurement
theory. We show that the L\"{u}ders probability of consecutive measurements
is a transition probability between two quantum states and that this probability 
cannot be treated as a quantum extension of the classical conditional probability. 
The Wigner distribution is shown to be a weighted transition probability that 
cannot be accepted as a quantum extension of the classical joint probability. 
We suggest the definition of quantum joint probabilities by introducing composite 
events in multichannel measurements. The notion of measurements under uncertainty
is defined. We demonstrate that the necessary condition for the mode interference 
is the entanglement of the composite prospect together with the entanglement of
the composite statistical state. As an illustration, we consider an example of  
a quantum game. A special attention is payed to the application of the approach 
to systems with multi-mode states, such as atoms, molecules, quantum dots, or
trapped Bose-condensed atoms with several coherent modes.  

\end{abstract}

\vskip 3mm

PACS numbers: 03.65.Ta, 02.50.Le, 03.67.Bg

\newpage

\section{Introduction}

The notion of quantum probabilities is a necessary ingredient of quantum
theory, which is of principal importance for the theory of quantum
measurements and quantum decision theory involving quantum information
processing \cite{Nielsen_1,Keyl_2,Galindo_3}. This notion has appeared
together with the arising quantum mechanics in the form of the Born rule
\cite{Born_4} defining the probabilities of measuring the eigenvalues of an
observable. The measurement procedure for a single observable is well
understood, being based on a projection-valued measure \cite{Reed_5}. The
consecutive measurement of two or more observables has been considered by
von Neumann \cite{Neumann_6} for the case of non-degenerate spectra and
generalized by L\"{u}ders \cite{Luders_7} for arbitrary spectra, including
the degenerate case. The probability of consecutive measurements, prescribed
by the L\"{u}ders rule, is often interpreted as a quantum extension of
classical conditional probability. Respectively, the Wigner distribution
\cite{Wigner_8}, that is, the weighted L\"{u}ders probability, is interpreted
as a quantum extension of classical joint probability. However, as is well
known, the quantum joint probabilities for two observables on a Hilbert space
can be mathematically correctly introduced only for compatible, that is, for
commuting observables
\cite{Neumann_6,Nelson_9,Gudder_10,Fine_11,Hall_12,Gudder_13,Malley_14,Malley_15,Malley_16}
or for observables from the Jordan algebra, where the
product of two operators is given by the symmetric Jordan form
\cite{Niestegge_17,Niestegge_18}. Such probabilities for commuting observables
or for the symmetric Jordan form enjoy the same properties as classical
probabilities. But the quantum joint probability for incompatible observables
remains undefined.

To be an extension of classical probability, quantum probability must satisfy
the {\it correspondence principle}, which was first advanced by Bohr
\cite{Bohr_19}, when analyzing atomic spectra. The principle requires that
quantum theory be reducible to classical theory in the limit where quantum
effects become negligible. In its general formulation, the quantum-classical
correspondence principle is understood as the requirement that the results of
quantum measurements would be reducible to those of classical measurements
when the quantum effects, such as interference, vanish. This reduction is
called decoherence \cite{Wheeler_20,Zurek_21}. In particular, for compatible
observables, quantum probability should reduce to a classical or quasiclassical
form.

In the present article, we analyze the problem of defining quantum probabilities
for arbitrary observables, whether compatible or incompatible. In order to give
precise results, we consider the most important case, when the spectra of
observables are non-degenerate. This case is of importance since then we have
one-to-one correspondence between the measured operator eigenvalues and the
related eigenfunctions, which makes straightforward the definition of all
probabilities, without the need of specifying the type of degeneracy, if it
would be present. At the same time, in practice, this situation does not hamper
the generality of the results, since it is always possible to resort to
von Neumann recipe \cite{Neumann_6} by slightly shifting the considered operator
so as to lift the degeneracy and to remove this shift at the end of
calculations.

Our main results are as follows:

(i) The L\"{u}ders probability of consecutive measurements is a transition
probability between two quantum states. It is symmetric with respect to events,
contrary to the generally asymmetric classical conditional probability. For
compatible events, the L\"{u}ders probability trivializes to the Kroneker delta.
The L\"{u}ders probability cannot be accepted as a quantum extension generalizing
the classical conditional probability.

(ii) The Wigner distribution is a weighted L\"{u}ders probability, that is, the
weighted transition probability. For compatible events, it trivializes to the
equality of both event probabilities. This distribution cannot be treated as a
quantum extension of the classical joint probability.

(iii) Quantum joint probabilities can be introduced as probabilities of composite
events represented by tensor products of events in two measurement channels. This
definition is valid for any event, including those corresponding to the
measurement of incompatible observables. Having in hands the general definition for
the joint probability makes it straightforward to define by the Bayes rule the
related conditional probability.

(iv) The probability of measurements under uncertainty is defined by employing
the positive operator-valued measure.

(v) It is shown that the mode interference can occur only for measurements performed 
under uncertainty, corresponding to entangled prospects, if the system state is also 
entangled. However entanglement is a necessary, but not sufficient, condition for the
mode interference.

(vi) The approach is illustrated by a quantum game demonstrating the existence of 
spontaneous breaking of average interference symmetry.

(vii) We show how the approach can be applied to multimode quantum systems subject 
to measurements under uncertainty. As examples of multi-mode systems, we keep in mind
such finite quantum systems with discrete spectrum as resonance atoms, molecules,
quantum dots, or trapped Bose-condensed atoms with several coherent modes.

\section{Algebra of quantum events}

First of all, it is necessary to recall the basic terminology that will be used
throughout the article. In different branches of science, one may employ different
terms for the same action, such as {\it accomplishing a measurement} or
{\it measuring an outcome}, used in the theory of measurements, or
{\it registering an effect}, often employed in information theory, or
{\it making a decision}, in decision theory, or {\it stating a proposition},
in logic, or just {\it observing an event}, which is customary for probability
theory. In what follows, we shall mostly use the terms {\it measuring an outcome}
and {\it observing an event}, implying all other synonymous meanings depending
on applications.

The set of events will be denoted as ${\cal R} = \{A_i: i = 1,2,\ldots\}$. The
set can be finite or infinite. While we shall explicitly deal with discrete sets
of events, the consideration can be straightforwardly generalized to continuous
sets. Events are connected by the rules of quantum logic \cite{Birkhoff_22}.
There exists the binary relation {\it addition}, or {\it disjunction}, or {\it union},
so that for any events $A, B \in {\cal R}$ there is $A \bigcup B \in {\cal R}$
meaning either $A$ or $B$. The addition is commutative, such that
$A \bigcup B = B \bigcup A$, associative,
$A \bigcup (B \bigcup C) = (A \bigcup B) \bigcup C$, and idempotent, $A \bigcup A = A$.

The other relation is {\it multiplication}, or {\it conjunction}, or {\it intersection},
such that for any $A, B \in {\cal R}$ there is $A \bigcap B \in {\cal R}$ meaning both $A$
and $B$. The multiplication is associative, so that
$A \bigcap B \bigcap C = (A \bigcap B) \bigcap C = A \bigcap (B \bigcap C)$,
and idempotent, $A \bigcap A = A$. Generally, it is not commutative,
$A \bigcap B \neq B \bigcap A$, and not distributive, in the sense
that $A \bigcap (B \bigcup C) \neq (A \bigcap B) \bigcup (A \bigcap C)$.

The set ${\cal R}$ includes the identical event $1$, which is, an event that
is identically true. For this event, $A \bigcap 1 = 1 \bigcap A = A$ and
$A \bigcup 1 = 1$, in particular, $1 \bigcup 1 = 1$. There also exists an
impossible event $0 \in {\cal R}$, which is identically false, so that
$A \bigcap 0 = 0 \bigcap A = 0$ and $A \bigcup 0 = A$, in particular,
$0 \bigcup 1 = 1$. The events for which $A \bigcap B = B \bigcap A = 0$
are called {\it disjoint} or {\it orthogonal}.

For each event $A \in {\cal R}$, there exists a {\it complementary}, or
negating, event $\bar{A} \in {\cal R}$, for which $A \bigcup \bar{A} = 1$ and
$A \bigcap \bar{A} = \bar{A} \bigcap A = 0$, in particular, $\bar{0} = 1$
and $\bar{1} = 0$.

The event set ${\cal R}$, with the above properties forms a non-commutative
non-distributive {\it event ring}. In quantum theory, events are associated
with the measurements of observables represented by self-adjoint operators
that do not necessarily commute. Non-commuting observables are called
incompatible, while those commuting are termed compatible. The non-distributivity
of quantum events can be illustrated by the Birkhoff-von Neumann example
\cite{Birkhoff_22}, defining three nontrivial events $A, B_1$ and $B_2$,
such that $B_1 \bigcup B_2 = 1$ and $A \bigcap B_1 = A \bigcap B_2 = 0$.
Then $A \bigcap (B_1 \bigcup B_2) = A \bigcap 1 = A$, however
$(A \bigcap B_1) \bigcup (A \bigcap B_2) = 0$.

The nonempty collection of all subsets of the event ring ${\cal R}$ including
${\cal R}$, which is closed with respect to countable unions and complements,
is the event sigma algebra $\Sigma$. The {\it algebra of quantum events} is the
pair $\{\Sigma, {\cal R}\}$ of the sigma algebra $\Sigma$ over the event
ring ${\cal R}$.

In practical problems, one considers not the whole event ring, but selected
events $\pi_j$ from ${\cal R}$, called prospects, which are assumed to form
a {\it prospect lattice}
$$
 \cL = \{ \pi_j : \; j = 1,2, \ldots \} \;  .
$$
The prospects of the lattice are ordered by means of the prospect
probabilities $p(\pi_j)$, such that
$$
 p(\pi_i) \leq p(\pi_j) \qquad (\pi_i \leq \pi_j) \;   .
$$
A prospect can represent a measurement, a proposition, or some event, whose
ordering can be done by defining the corresponding probabilities. The number
of prospects in the lattice ${\cal L}$ can be finite or infinite. The main
problem is how to correctly define the probabilities for quantum events.

\section{Probability of separate events}

The way of characterizing the probabilities of separate events, representing
quantum measurements, is well known, being based on a projection-valued measure
\cite{Reed_5}. Von Neumann \cite{Neumann_6} mentioned that the measurement
procedure is equivalent to decision making. In the literature, the theory of
quantum measurements is often classified as decision theory
\cite{Benioff_23,Holevo_24}.

A quantum system is described by a Hilbert space of microstates
${\cal H}$ and a statistical operator, or system state $\hat{\rho}(t)$ on
${\cal H}$, which generally can be a function of time $t \geq 0$. The system
state is a non-negative operator normalized to one, ${\rm Tr} \hat{\rho}(t) = 1$,
with the trace over ${\cal H}$. The pair $\{{\cal H}, \hat{\rho}(t)\}$ is termed
{\it quantum statistical ensemble}. Observable quantities are represented by
self-adjoint operators $\hat{A}$ on ${\cal H}$ forming an algebra of local
observables ${\cal A} \equiv \{\hat{A}\}$. Observable quantities are given by
the operator expected values
\be
\label{1}
 \lgl \hat A(t) \rgl \equiv {\rm Tr}\hat\rho(t) \hat A \;  ,
\ee
where the trace is over ${\cal H}$. For generality, we shall be using mixed
system states, whose particular case is a pure state
$\hat{\rho} = |\psi \rangle \langle \psi|$. Actually, a real system cannot
be completely isolated from its surrounding, but always experiences its
influence. There can exist only quasi-isolated systems
\cite{Yukalov_25,Yukalov_26,Yukalov_27}. Hence, real systems, generally,
have to be described by mixed states. A quantum system is a set
$\{{\cal H}, \hat{\rho}(t), {\cal A}\}$ of the statistical ensemble and the
algebra of local observables.

The operators of observables, being self-adjoint, possess real-valued
eigenvalues $A_n$, labeled by a multi-index $n$ and given by the eigenproblem
\be
\label{2}
 \hat A | n \rgl = A_n | n \rgl \;  ,
\ee
with the eigenfunctions $|n\rangle$ forming a complete orthonormal basis
$\{|n\rangle\}$. The Hilbert space ${\cal H}$ can be defined as the closed
linear envelope ${\cal H} = {\rm span} \{|n\rangle\}$.

The operator spectrum $\{A_n\}$ can be discrete or continuous, degenerate or
non-degenerate. For concreteness, we write below the formulas as corresponding
to discrete spectra. This will make it clearer the principal points we aim
at discussing, without complications related to continuous spectra. The extension
to the latter is straightforward. Moreover, in many cases, it is possible to avoid
continuous spectra by imposing appropriate boundary conditions. For instance, the
standard procedure of dealing with discrete momenta is by considering a quantum
system in a finite volume. The passage to continuous momenta is commonly done by
taking the thermodynamic limit.

In order to avoid degenerate spectra, it is possible, as has been suggested by
von Neumann \cite{Neumann_6}, to lift the degeneracy by slightly shifting the operator
with a small term breaking this degeneracy, and sending the additional term to zero
at the end of calculations. This procedure is somewhat similar to the Bogolubov
method of symmetry breaking by introducing infinitesimal terms
\cite{Bogolubov_28,Bogolubov_29}.

With the eigenfunctions $|n\rangle$, one can introduce the projection operators
\be
\label{3}
 \hat P_n \equiv | n \rgl \lgl n |
\ee
that are self-adjoint idempotent operators, such that
${\hat P}^2_n = {\hat P}_n = {\hat P}_n^+$. The projectors are orthogonal
and provide the resolution of unity,
\be
\label{4}
  \hat P_m \hat P_n = \dlt_{mn} \hat P_n \; , \qquad
\sum_n \hat P_n = \hat 1_\cH \; ,
\ee
where ${\hat 1}_{\cal H}$ is the identity operator in ${\cal H}$. The operators
of observables enjoy the spectral decomposition
\be
\label{5}
 \hat A = \sum_n A_n \hat P_n \;  ,
\ee
where the summation is over the total set $\{n\}$ of multi-indices.

Performing measurements of an observable ${\hat A}$, one can get one of the
eigenvalues $A_n$. Denoting the prospects of finding $A_n$ by the same letter
$A_n$ as the related eigenvalue, we have the prospect lattice
${\cal L} = \{A_n\}$. Assuming, for simplicity, a nondegenerate spectrum, one
has the {\it correspondence}
\be
\label{6}
 A_n ~ \ra ~ |n \rgl ~ \ra ~ \hat P_n \;  .
\ee
A projector ${\hat P}_n$ represents a proposition, thus, the set
${\cal P} \equiv \{{\hat P}_n\}$ is a {\it proposition lattice} isomorphic to
the prospect lattice ${\cal L}$. Because of the properties of the projectors,
their set $\{{\hat P}_n\}$ forms an operator probability measure that is an
orthogonal projection measure. The triple
$\{{\cal H}, {\cal P}, \hat{\rho}(t)\}$ is the {\it quantum probability space}.

According to the Gleason theorem \cite{Gleason_30}, for a Hilbert space of
dimension larger than two, the only possible measure for the probability of
measuring $A_n$, in the system state $\hat{\rho}(t)$, must have the form
\be
\label{7}
 p(A_n,t) \equiv {\rm Tr}\hat\rho(t) \hat P_n \;  ,
\ee
with the properties
$$
  \sum_n p(A_n,t) = 1 \; , \qquad 0 \leq p(A_n,t) \leq 1 \; .
$$
For a measurement at $t=0$, we shall write $\hat{\rho} \equiv \hat{\rho}(0)$.
Then the probability of $A_n$ becomes
\be
\label{8}
 p(A_n) \equiv p(A_n,0) = {\rm Tr}\hat\rho \hat P_n \;  ,
\ee
which, with notation (1), is the average
\be
\label{9}
 p(A_n)  = \lgl \hat P_n \rgl \;  .
\ee
The family $\{p(A_n)\}$ forms a probability measure.

Taking the trace over the basis of the eigenfunctions $|n\rangle$ leads to
\be
\label{10}
 p(A_n)  = {\rm Tr}\hat\rho \hat P_n = \lgl n \;|\; \hat\rho \; | \; n \rgl \;  .
\ee
The expected value of an observable $\hat{A}$ reads as
\be
\label{11}
 \lgl \hat A \rgl = \sum_n p(A_n) A_n \;  .
\ee
In the case of a pure system state, $p(A_n) = |\langle n|\psi\rangle|^2$.
The most probable prospect $A_*$ is given by the condition
\be
\label{12}
 p(A_*) \equiv \sup_n p(A_n) \;  .
\ee

In addition to the probability of separate events, it is possible to define
the probability of the union of disjoint events, such that
$A_m \bigcap A_n = \delta_{mn}$, for which the related projectors are
orthogonal, ${\hat P}_m {\hat P}_n = 0$, where $m \neq n$. Then the union
$A_m \bigcup A_n$ is represented as ${\hat P}_m + {\hat P}_n$. Therefore,
\be
\label{13}
 p\left (A_m\bigcup A_n \right ) =
{\rm Tr}\hat\rho \left (\hat P_m + \hat P_n \right ) =
p(A_m) + p(A_n) \;  ,
\ee
for $m \neq n$ and $A_m \bigcap A_n = 0$. The generalization to an arbitrary
number of mutually disjoint events is straightforward.

But it is important to stress that the summation formula
\be
\label{14}
p\left ( A\bigcup B \right ) = \lgl \hat P_A + \hat P_B \rgl = p(A) + p(B)
\ee
is valid if and only if the events $A$ and $B$ are {\it disjoint}, such that
\be
\label{15}
A \bigcap B = 0 \; , \qquad {\hat P}_A {\hat P}_B = 0.
\ee
It is easy to show that formula (14) does not work for not disjoint events.
For instance, let us consider the sum $A \bigcup A = A$, with $A \neq 0$.
If one would use the summation formula for the above equality, then the
left-hand side of this equality $A \bigcup A = A$ would give $2 p(A)$, while
the right-hand side yields $p(A)$, which is meaningless.

Also, for non-disjoint events, one cannot use the classical relation
$p(A \bigcup B) = p(A) + p(B) - p(A \bigcap B)$, since the quantum joint
probability for incompatible observables is not defined. The Kirkwood
\cite{Kirkwood_31} form $\langle {\hat P}_A {\hat P}_B \rangle$ does not
constitute a probability, being complex for incompatible observables.

\section{Quantum state reduction}

Measurements influence the system. Thus, performing the measurement of an
observable ${\hat A}$, and getting $A_m$ as an outcome, implies that
the system state ${\hat \rho}$ has been changed by the measurement to ${\hat \rho}'$,
such that $p(A_m) = \langle m|{\hat \rho}'|m \rangle = 1$ and all other probabilities
$p(A_n)$, with $m \neq n$, are zero. That is, the matrix element
$\langle n|{\hat \rho}'|n \rangle$ is equal to $\delta_{mn}$. In the case
of a nondegenerate spectrum considered by von Neumann \cite{Neumann_6},
this means that the state reduction ${\hat \rho} \ra {\hat P}_m$ has occurred. L\"{u}ders
\cite{Luders_7} generalized the consideration for an arbitrary spectrum,
including degenerate ones, so that the state reduction takes the form
\be
\label{16}
 \hat\rho ~ \longrightarrow ~
\frac{\hat P_m \hat\rho \hat P_m}{{\rm Tr}\hat\rho\hat P_m} \;  .
\ee
For a pure system state, one has the reduction
$|\langle n| \psi \rangle| \ra \delta_{mn}$. And one says that the wave
function $\psi$ collapses to $|m \rangle$.

There have been numerous discussions of what the state reduction could mean,
whether it is a discontinuous jump in an objective system state, or the
system state is a subjective construct of an observer. In this later interpretation,
the subjective construct would characterize the belief
propagation, but not an objective property of the physical system
\cite{Caves_32,Leifer_33,Leifer_34,Fuchs_35}. Accordingly, the state reduction would be
just a Bayesian update of information in the mind of the observer. A good
discussion of dynamical versus inferential conceptions in quantum measurements
has been recently done by Wallace \cite{Wallace_35}.

The state reduction does not need to be interpreted as a sudden collapse. It
only looks like that, when one neglects the existence of a measurement
procedure, treating the latter as an instantaneous receipt of information. Any
real measurement requires finite time and involves interactions with measuring
devices and observers \cite{Wallace_35,Neumann_6,Margenau_36,Moladuer_37,Ludwig_38}.
Even the so-called nondemolition and nondestructive measurements may essentially
influence the measured system \cite{Braginsky_39,Merkli_40,Merkli_41,Yukalov_42}.
In what follows, the environment, including measuring apparatuses and observers,
acting on the system in the process of measurement, will be called for short
a {\it measurer}.

Let the Hilbert space describing the system microscopic states be denoted by
${\cal H}_S$ and the Hilbert space of the measurer, by ${\cal H}_M$. The complex
object, composed of the measured system and the measurer is characterized by the
Hilbert space
\be
\label{17}
 \cH_{SM} = \cH_S \bigotimes \cH_M \; .
\ee
The corresponding statistical ensemble is the pair
$\{{\cal H}_{SM}, {\hat \rho}_{SM}(t)\}$. At the initial time $t = 0$, if the
system is in a state ${\hat \rho}_{SM}(0)$, then, during the process of
measurement, the state changes to ${\hat \rho}_{SM}(t)$. By the Kadison theorem
\cite{Kadison_43}, there exists a one-parameter family of unitary operators
${\hat U}_{SM}(t)$, such that
\be
\label{18}
 \hat\rho_{SM}(t) = \hat U_{SM}(t) \hat\rho_{SM}(0) \hat U_{SM}^+(t) \;  .
\ee
The evolution operators ${\hat U}_{SM}(t)$ characterize the quantum dynamics
of the complex system \cite{Polkovnikov_44}.

Suppose one performs the measurement of an observable ${\hat A}$ defined
on the system space ${\cal H}_S$. An event $A_n$ is represented by the projector
${\hat P}_n$. The probability of measuring $A_n$, at time $t$, is
\be
\label{19}
 p(A_n,t) \equiv
{\rm Tr}_{SM} \hat\rho_{SM}(t) \hat P_n \bigotimes \hat 1_M \;  ,
\ee
where ${\hat 1}_M$ is the identity operator in ${\cal H}_M$ and the trace is
over ${\cal H}_{SM}$. Equation (19) can be rewritten as
\be
\label{20}
  p(A_n,t) = {\rm Tr}_{S} \hat\rho_{S}(t) \hat P_n  \; ,
\ee
with the trace over ${\cal H}_S$ and the reduced statistical operator
\be
\label{21}
   \hat\rho_{S}(t) \equiv  {\rm Tr}_{M} \hat\rho_{SM}(t)\; ,
\ee
where the trace is over ${\cal H}_M$. Note that the evolution of the
reduced state (21), generally, is not described by a unitary operator.
Using the basis of the eigenfunctions of ${\hat A}$ yields
\be
\label{22}
 p(A_n,t) = \lgl n \; |\; \hat\rho_S(t) \; |\; n \rgl \;  .
\ee

The change of the system state is caused by the interactions with the measurer.
If at time $t_m$, one finds a value $A_m$, so that $p(A_m,t_m) = 1$, this means
that
\be
\label{23}
  \lgl n \; |\;  {\rm Tr}_{M} \hat\rho_{SM}(t_m)\; |\; n \rgl =
\dlt_{mn} \; .
\ee
In other words, the interactions with the measurer have transformed the
initial state of the combined system plus the measurer ${\hat \rho}_{SM}(0)$
into the final state ${\hat \rho}_{SM}(t_m)$, for which one could write
\be
\label{24}
 \hat\rho_S(t_m) = {\rm Tr}_{M} \hat\rho_{SM}(t_m) =
\frac{\hat P_m\hat\rho_S(0)\hat P_m}{{\rm Tr}_S \hat\rho_S(0) \hat P_m} \;,
\ee
at this given time $t_m$. In that sense, there is no any sudden collapse,
but there is a gradual transformation due to interactions during a finite
time $t_m$.

The so-called quantum state collapse arises only when one treats a measurement
procedure not as a real interaction of the system with a measurer during a
finite time, but as an imaginary process of instantaneously receiving information.
There is no collapse under realistic measurements. However, assuming that
the time of measurement is short, it is admissible, for convenience, to formally
consider the limit $t_m \ra 0$. Then, one can deal with the probability
\be
\label{25}
 p(A_n) \equiv \lim_{t\ra+0} p(A_n,t) \;  ,
\ee
keeping in mind that this is just a convenient way of consideration {\it for
all practical purposes} \cite{Bell_45}. Concrete models describing the
dynamics of quantum measurements have been given, e.g., in Refs.
\cite{Barchielli_45,Ghirardi_45,Pearle_45}.

\section{Probability of consecutive measurements}

Two observables, say ${\hat A}$ and ${\hat B}$, defined on the same Hilbert
space, even if they do not commute with each other, can be measured consecutively.
Therefore, one often considers such consecutive measurements as a possible way
allowing for the introduction of quantum joint and conditional probabilities
generalizing the related classical notions. In the present section, we show that
such probabilities cannot be treated as extensions of the corresponding classical
notions.

Let us assume that one first measures the observable ${\hat B}$ on ${\cal H}$ in
a state ${\hat \rho}$. The eigenproblem
\be
\label{26}
  \hat B | \al \rgl = B_\al | \al \rgl
\ee
makes it possible to define the correspondence
\be
\label{27}
  B_\al ~ \ra ~ |\al\rgl ~ \ra ~ \hat P_\al \; ,
\ee
where ${\hat P}_\alpha \equiv |\alpha \rangle \langle \alpha|$ is a projection
operator. Again, for simplicity, we assume a nondegenerate spectrum, which is
not essential but just makes the consideration more clear and persuasive. Recall,
that to avoid technical complications of dealing with degenerate spectra, it is
always possible to use the von Neumann recipe of slightly shifting the considered
operator, lifting by this the degeneracy \cite{Neumann_6}. The family of the
eigenfunctions $|\alpha \rangle$ forms a basis in ${\cal H}$, and the observable
${\hat B}$ can be written as
\be
\label{28}
\hat B = \sum_\al B_\al \hat P_\al \;   .
\ee

Suppose that the measurement gives $B_\alpha$, which implies, after the measurement,
that $p(B_\alpha) = 1$ and $p(B_\beta) = \langle \beta |{\hat \rho}| \beta \rangle$
equals $\delta_{\alpha \beta}$. This means that the state ${\hat \rho}$ reduces to
\be
\label{29}
 \hat\rho_\al \equiv
\frac{\hat P_\al\hat\rho\hat P_\al}{{\rm Tr}\hat\rho\hat P_\al} \;  .
\ee
Here we keep in mind the explanation of the state reduction given in the previous
section, as due to the actual measurement procedure, when assuming the limiting
case (25).

Immediately after this first measurement, we measure another observable ${\hat A}$,
acting on the same Hilbert space ${\cal H}$. This observable, generally, does not
commute with ${\hat B}$. Hence the states $|n \rangle$ and $|\alpha \rangle$ are not
necessarily orthogonal, so the projectors ${\hat P}_n$ and ${\hat P}_\alpha$ are not
orthogonal as well. The measurement of $\hat A$, in state (29), results in the event
$A_n$ with the probability
\be
\label{30}
 p_L(A_n|B_\al) \equiv {\rm Tr}\hat\rho_\al \hat P_n \;  ,
\ee
which is called the {\it L\"{u}ders transition probability}. Here the trace is over
${\cal H}$. Employing definition (29) and taking account of the equality
$$
 {\rm Tr} \hat\rho \hat P_\al  = \lgl \hat P_\al \rgl = p(B_\al)
$$
yields
\be
\label{31}
  p_L(A_n|B_\al) = \frac{{\rm Tr}\hat\rho\hat P_\al\hat P_n\hat P_\al}
{{\rm Tr}\hat\rho\hat P_\al} =
\frac{\lgl \hat P_\al\hat P_n\hat P_\al\rgl}{\lgl \hat P_\al\rgl}\; .
\ee

The numerator in Eq. (31) is the form introduced by Wigner \cite{Wigner_8},
because of which it is usually called \cite{Johansen_46} the {\it Wigner
distribution}, which is
\be
\label{32}
 p_W(A_n|B_\al)  \equiv  {\rm Tr}\hat\rho\hat P_\al\hat P_n\hat P_\al =
\lgl  \hat P_\al\hat P_n\hat P_\al \rgl \; .
\ee
Hence, the L\"{u}ders transition probability (31) takes the form
$$
p_L(A_n|B_\al) = \frac{p_W(A_n|B_\al)}{p(B_\al)} \;   ,
$$
which defines the relation
\be
\label{33}
 p_W(A_n|B_\al) = p_L(A_n|B_\al) p(B_\al) \;  .
\ee
This relation looks similarly to the relation between the joint and conditional
probabilities in classical probability theory. Because of this, the temptation
arises to treat the L\"{u}ders probability $p_L(A_n|B_\alpha)$ as a quantum
extension of the classical conditional probability and the Wigner distribution
$p_W(A_n|B_\alpha)$ as a quantum extension of the classical joint probability.

However, a closer look proves that the formal analogy here is misleading.
First, we may notice that the product ${\hat P}_\alpha {\hat P}_n {\hat P}_\alpha$
is not a projector for incompatible observables. Then, taking into account
the equality
$$
\lgl n \; |\; \hat P_\al \;|\; n \rgl = \lgl \al | \hat P_n | \al \rgl =
| \lgl \al \; | \; n \rgl |^2 \;  ,
$$
we see that the Wigner distribution reads as
\be
\label{34}
 p_W(A_n|B_\al) = | \lgl n \; | \; \al \rgl |^2 p(B_\al) \;  ,
\ee
while the L\"{u}ders probability is
\be
\label{35}
 p_L(A_n|B_\al) =   {\rm Tr}\hat P_n\hat P_\al =
| \lgl n \; | \; \al \rgl |^2 \;  .
\ee
The latter is nothing but a transition probability between two quantum states,
$|n \rangle$ and $|\alpha \rangle$, with the following standard properties
for the transition probability
$$
\sum_n p_L(A_n|B_\al) =  \sum_\al p_L(A_n|B_\al) = 1 \; , \qquad
 p_L(A_n|A_n) =  p_L(B_\al|B_\al) = 1 \; .
$$
And the Wigner distribution is a weighted transition probability.

The most important point is that the L\"{u}ders transition probability is always
symmetric, such that
\be
\label{36}
  p_L(A_n|B_\al) = p_L(B_\al|A_n) \; ,
\ee
for arbitrary events, whether compatible or incompatible. In contrast, the classical
conditional probability, generally, is not symmetric. Hence the L\"{u}ders
transition probability, generally, cannot be reduced to the classical
conditional probability. Moreover, for compatible observables, instead of
leading to meaningful classical counterparts, both the L\"{u}ders transition
probability and the Wigner distribution become trivial:
$$
 p_L(A_n|B_\al) ~ \ra \dlt_{n\al} \; , \qquad
p_W(A_n|B_\al) ~ \ra \dlt_{n\al} p(B_\al) = \dlt_{\al n} p(A_n) \;  .
$$

Thus, we come to the conclusion that the L\"{u}ders probability is a
transition probability that cannot be treated as a quantum extension of the
classical conditional probability. Respectively, the Wigner distribution
is a weighted transition probability that cannot be considered as a quantum
extension of the classical joint probability.

Recall that the Kirkwood \cite{Kirkwood_31} form
$$
\lgl \hat P_n \hat P_\al \rgl = \sum_m
\lgl m \;|\; \hat\rho \;|\; n \rgl \lgl n \;|\; \al \rgl \lgl \al\; |\; m \rgl
$$
also cannot be accepted as a joint quantum probability, since it is complex
for incompatible observables, while for compatible observables it
trivializes to
$$
 \lgl \hat P_n \hat P_\al \rgl ~ \ra ~ \dlt_{n\al} p(B_\al) =
\dlt_{\al n} p(A_n) \;  .
$$

It is worth emphasizing that, in addition to relation (33), it is
straightforward to derive several other formal identities that, however, do
not necessarily enjoy the meaning of equations generalizing the corresponding
relations for classical probabilities. For instance, using the projector
expansions
$$
 \sum_n \hat P_n = \sum_\al \hat P_\al = \hat 1_\cH
$$
and the identities
$$
   p(A_n) \equiv  \lgl \hat P_n \rgl \equiv
\lgl \hat P_n \hat 1_\cH \rgl \equiv
\lgl \hat 1_\cH \hat P_n \hat 1_\cH \rgl \;  ,
$$
it is possible to produce an infinite chain of other identities
$$
\lgl \hat P_n \rgl \equiv \sum_\al \lgl \hat P_n \hat P_\al \rgl
\equiv \sum_{\al\bt} \lgl \hat P_\al\hat P_n \hat P_\bt \rgl
\equiv \sum_{\al\bt}
\sum_m \lgl \hat P_m \hat P_\al \hat P_n \hat P_\bt \rgl \equiv
$$
\be
\label{37}
 \equiv \sum_{\al\bt}
\sum_{mk} \lgl \hat P_m \hat P_\al \hat P_n \hat P_\bt \hat P_k \rgl \; ,
\ee
and so on. The meaning of this chain is that the measurement of $A_n$ can
be done through an arbitrary sequence of other measurements. But the summed
terms in chain (37) are not probabilities.

One should not confuse identities and changes of notations with meaningful definitions.
For example, assuming the validity of $\bigcup_\alpha B_\alpha = 1$, one can
get the equalities
$$
 A_n = A_n \bigcap 1 = A_n \bigcap \left ( \bigcup_\al B_\al \right ) \; ,
$$
and many others like that. This makes it admissible to get the identities
\be
\label{38}
  p(A_n) = p\left ( A_n \bigcap 1 \right ) =
p\left (A_n \bigcap \left ( \bigcup_\al B_\al \right ) \right ) \; .
\ee
These identities, however, must not be treated as a definition of the joint
probability. Their meaning is nothing but the trivial sequence of the
identities $p(A_n) \equiv p(A_n) \equiv p(A_n)$ expressed in different forms. Combining
identities (37) and (38), one can produce a number of other identities having the
same meaning as the definition $p(A_n) \equiv \langle {\hat P}_n \rangle$, but just
rewritten in different forms. Thus, one can write
\be
\label{39}
 p\left ( A_n \bigcap \left ( \bigcup_\al B_\al \right )\right ) =
\sum_\al \lgl \hat P_n \hat P_\al \rgl \; .
\ee
Again, this is not a definition of the joint probability, but just a
rewriting of the definition $p(A_n) \equiv \langle {\hat P}_n \rangle$.
It would be wrong to interpret the expression
$A_n \bigcap (\bigcup_\alpha B_\alpha)$ as corresponding to the sum
$\sum_\alpha {\hat P}_n {\hat P}_\alpha$, because
the joint probability of quantum events has not been defined and because,
in quantum logic, the event ring is not distributive and thus
$A_n \bigcap (\bigcup_\alpha B_\alpha)$ does not equal
$\bigcup_\alpha (A_n \bigcap B_\alpha)$. Also, the right-hand side of
Eq. (39) is the sum of the Kirkwood forms that are not probabilities.

By employing Eqs. (37) and (38), it is also possible to produce the
formal relation
\be
\label{40}
 p\left (A_n \bigcap \left ( \bigcup_\al B_\al \right ) \right ) =
\sum_\al p_W(A_n|B_\al) +
\sum_{\al\neq\bt} \lgl \hat P_\al \hat P_n \hat P_\bt \rgl \;   .
\ee
The temptation arises to treat here the left-hand side as a total joint
probability, while the first term in the right-hand side is viewed as a sum of
partial joint probabilities and the last double sum is seen as an interference
term between different events. In this interpretation, relation (40) would be
assumed to be a quantum generalization of the classical summation formula for
the total joint probability expressed through the sum of partial joint probabilities.
Such an interpretation is widespread in many applications of quantum information
processing to decision theory \cite{Busemeyer_47}. But this interpretation is
principally wrong for several reasons. First, the quantum joint probability
for observables on the same Hilbert space is not defined. Second, the Wigner
distribution, as explained above, is just a weighted transition probability
and cannot be accepted as a quantum extension of the classical joint probability.
Third, relation (40) does not satisfy the correspondence principle, according to
which, for compatible observables, relation (40) should become the classical
summation formula for the total joint probability. Really, for compatible
observables, Eq. (40) becomes the trivial identity $p(A_n) = p(A_n)$. This is not
surprising, since relation (40) has been derived as a rewriting
of the definition $p(A_n) \equiv \langle {\hat P}_n \rangle$. Therefore,
from the very beginning, relation (40) cannot reveal more information than the
identity $p(A_n) = p(A_n)$, which becomes explicit for compatible observables.

Rewriting an identity in different forms is not providing here new information
and, in particular, does not provide any clue on how to define joint quantum
probabilities for observables acting on the same Hilbert space.

\section{Probability of generalized propositions}

Real quantum measurements are treated as propositions providing a definitive
receipt for the evaluation of the event probabilities \cite{Neumann_6},
such as $p(A_n)$. These measurements are termed {\it operationally testable
propositions} \cite{Randall_48}. It is possible to consider a mathematically
more general set of quantum propositions with indefinite answers. For example,
one accomplishes a measurement of an observable ${\hat B}$, but the result of
the measurement is not explicitly known, that is, one is not sure which of
the eigenvalues $B_\alpha$ is obtained. Such a situation can be referred to by
different terms, e.g., an uncertain measurement, fuzzy measurement, not
completely defined measurement, inconclusive measurement, ambiguous measurement,
or generalized measurement \cite{Yuen_49,Huttner_50}. Formally, this case could
be related to non-classical logic in quantum mechanics \cite{Suppes_51,Putnam_52}.
The corresponding generalized propositions are not realized in operationally
tested measurements, but they serve as important tools at intermediate stages
of quantum information processing \cite{Nielsen_1,Keyl_2,Galindo_3}. In what
follows, we shall need the related mathematical constructions. For this,
we now introduce the main definitions to be used later.

Suppose we measure an observable ${\hat B}$, defined on a Hilbert space
${\cal H}$, with the set of eigenvalues $B \equiv \{B_\alpha\}$. The vector
\be
\label{41}
 |B \rgl = \sum_\al b_\al | \al \rgl
\ee
in quantum information processing is termed a {\it multimode state}. Vector (41)
does not need to be necessarily normalized to one. Varying the coefficients
$b_\alpha$ yields a manifold $\mathbb{B}$ of admissible sets $B$. Such multimode
states can currently be created in different experiments
\cite{Nielsen_1,Keyl_2,Galindo_3,Lupo_53,Shi_53,Tannor_53,Dobek_53}.
The generalized proposition operator is
\be
\label{42}
 \hat P_B \equiv | B \rgl \lgl B | \;  .
\ee
This operator is not necessarily a projector, since the multimode state can be
not normalized to one. But it is required that the resolution of unity be valid:
\be
\label{43}
 \sum_{B\in\mathbb{B}} \hat P_B = \hat 1_\cH \;  .
\ee
The family $\{{\hat P}_B: B \in \mathbb{B}\}$ of the generalized proposition
operators forms a {\it positive operator-valued measure}
\cite{Nielsen_1,Davies_54,Kraus_55,Holevo_56}.

Similarly to Eq. (26), one can consider the correspondence
\be
\label{44}
 B ~ \ra ~ | B \rgl ~ \ra ~ \hat P_B \;  .
\ee
Therefore the probability of a multimode state is
\be
\label{45}
 p(B) \equiv {\rm Tr}\hat\rho \hat P_B = \lgl \hat P_B \rgl \;  ,
\ee
which gives
\be
\label{46}
  p(B) =
\sum_{\al\bt} b_\al^* b_\bt \lgl \al \;|\; \hat\rho \;|\; \bt \rgl \;  .
\ee
By construction, the probabilities $p(B)$ compose a probability
measure with the standard properties
\be
\label{47}
 \sum_{B\in\mathbb{B}} p(B) = 1 \; , \qquad 0 \leq p(B) \leq 1 \;  .
\ee
Separating in Eq. (46) the terms with $\alpha = \beta$ and $\al\neq\bt$
yields
\be
\label{48}
  p(B) = \sum_\al |b_\al|^2 p(B_\al) + q(B),
\ee
where
\be
\label{49}
p(B_\al) \equiv \lgl \hat P_\al \rgl =
\lgl \al\; |\; \hat\rho \; |\; \al\rgl \; \qquad
q(B) = \sum_{\al\neq\bt} b_\al^* b_\bt \lgl \al \; |\; \hat\rho\; |\; \bt \rgl \; .
\ee

The generalized proposition operators ${\hat P}_B$ introduced above
characterize formal operational propositions \cite{Foulis_57,Foulis_58},
as compared to the projectors ${\hat P}_\alpha$ corresponding to
operationally testable events. The relation between the generalized
proposition operator and the projectors ${\hat P}_\alpha$ is as follows:
\be
\label{50}
 \hat P_B = \sum_\al |b_\al|^2 \hat P_\al +
\sum_{\al\neq\bt} b_\al b_\bt^* | \al \rgl \lgl \bt | \;  .
\ee
The positive operator-valued measure $\{{\hat P}_B: B \in \mathbb{B}\}$
is a generalization of the orthogonal projection measure $\{\hat P_\alpha\}$.
The later, being a particular case of $\{{\hat P}_B: B \in \mathbb{B}\}$,
is one of its filtrations.

As was stressed above, the multimode states (41) are often realized in
experiments. Therefore, expression (45) can be understood as the probability
of preparing such a multimode state. It can also be interpreted as the
probability of a non-destructive measurement of the initial state (41),
minimally disturbing the state \cite{Yukalov_59,Yukalov_60}, when the
measurement results merely in the appearance of factors $b_\al=e^{i\varphi}$,
with real random phases $\varphi$.

\section{Multichannel measurement procedure}

Incompatible observables, which do not commute with each other, cannot
be measured simultaneously. Their measurement requires to employ a more
complicated procedure. Such a general procedure, which can be used for
measuring any type of observables, whether compatible or not, can be
constructed as follows.

Suppose we need to measure two observables, ${\hat A}$ and ${\hat B}$,
which in general are not compatible. With the eigenfunctions of these
operators, $|n \rangle$ and $|\alpha \rangle$ respectively, one can
define two copies of the Hilbert space,
\be
\label{51}
 \cH_A \equiv {\rm span} \{ | n \rgl \} \; , \qquad
\cH_B \equiv {\rm span} \{ | \al \rgl \} \  .
\ee

Transitions between different system states, defined on different spaces,
can be characterized by involving the Neumark theorem \cite{Gelfand_61}
and the notion of Kraus operators \cite{Kraus_55,Kraus_62}. The
equivalent and physically transparent way is to consider quantum channels
representing completely positive linear mappings
\cite{Keyl_2,Holevo_63,Holevo_64}.

Assume that, at time $t = 0$, we are interested in the observable ${\hat B}$
defined on ${\cal H}_B$, with the system state being ${\hat \rho}_B(0)$
on ${\cal H}_B$. Starting the measurement, we connect the system with
a measurer in the state ${\hat \rho}_M(0)$ acting on ${\cal H}_M$, so that
the composite system state becomes
${\hat \rho}_B(0) \bigotimes {\hat \rho}_M(0)$ on the space
${\cal H}_{BM} \equiv {\cal H}_B \bigotimes {\cal H}_M$. The corresponding
channel is the mapping
\be
\label{52}
C_0 : \; \hat\rho_B(0) ~ \ra ~
 \hat\rho_B(0) \bigotimes \hat\rho_M(0) \;  .
\ee

The process of measurement requires some time during which the composite system
evolves to an entangled state
\be
\label{53}
 \hat\rho_{BM}(t) =
\hat U_{BM}(t) \hat\rho_B(0) \bigotimes \hat\rho_M(0) \hat U_{BM}^+(t) \;  .
\ee
The transition from $t = 0$ to time $t_1 > 0$ is given by the evolution channel
\be
\label{54}
 C_1 : \; \hat\rho_B(0) \bigotimes \hat\rho_M(0) ~ \ra ~
 \hat\rho_{BM}(t_1) \;  .
\ee

If the readout of the result is taken at time $t_2 > t_1$, this corresponds
to the disentangling channel
\be
\label{55}
  C_2 : \; \hat\rho_{BM}(t_1) ~\ra ~
\hat\rho_{B}(t_2) \bigotimes  \hat\rho_{M}(t_2) ,
\ee
where
\be
\label{56}
 \hat\rho_{B}(t_2) \equiv {\rm Tr}_M \hat\rho_{BM}(t_2) \; , \qquad
\hat\rho_{M}(t_2) \equiv {\rm Tr}_B \hat\rho_{BM}(t_2) \;  .
\ee

Continuing the measurement further entangles again the system state, leading
at $t > t_2$ to the state
\be
\label{57}
  \hat\rho_{BM}(t) = \hat U_{BM}(t-t_2) \hat\rho_B(t_2)
\bigotimes \hat\rho_M(t_2) \hat U_{BM}^+(t-t_2) \; .
\ee
The related transition for $t_3 > t_2$ is described by the channel
\be
\label{58}
 C_3 : \; \hat\rho_{B}(t_2) \bigotimes  \hat\rho_{M}(t_2) ~ \ra ~
 \hat\rho_{BM}(t_3) \;  .
\ee

In order to perform the measurement of the observable ${\hat A}$ defined on the
space ${\cal H}_A$, one needs to transform the basis $\{|\alpha \rangle \}$ (for
the observable ${\hat B}$) to the basis $\{|n \rangle \}$ (for the observable
${\hat A}$), which is realized by means of a unitary basis transformation
${\hat T}_{AB}$ connecting the copies ${\cal H}_B$ and ${\cal H}_A$. This is
equivalent to the state transformation
\be
\label{59}
 \hat\rho_{AM}(t) = \hat T_{AB} \hat U_{BM}(t-t_3)\hat\rho_{BM}(t_3)
\hat U_{BM}^+(t-t_3) \hat T_{AB}^+ \; ,
\ee
with the state $\hat{\rho}_{AM}$ acting on
${\cal H}_{AM} \equiv {\cal H}_{A} \bigotimes {\cal H}_{M}$. Such a procedure of
preparing the measurer for another measurement at $t_4 > t_3$ is characterized by
the channel
\be
\label{60}
 C_4 : \; \hat\rho_{BM}(t_3)  ~ \ra ~ \hat\rho_{AM}(t_4) \; .
\ee

The readout of the result for the observable ${\hat A}$, at $t_5 > t_4$,
implies the reduction to the state
\be
\label{61}
  \hat\rho_A(t_5) = {\rm Tr}_M \hat\rho_{AM}(t_5) \; ,
\ee
acting on ${\cal H}_{A}$, which is given by the channel
\be
\label{62}
 C_5 : \; \hat\rho_{AM}(t_4)  ~ \ra ~ \hat\rho_{A}(t_5) \;  .
\ee

The whole described procedure is the convolution of the channels
$$
C_0 \bigotimes C_1  \bigotimes C_2 \bigotimes C_3  \bigotimes C_4 \bigotimes C_5 : \;
\hat\rho_B ~ \ra ~ \hat\rho_B\bigotimes\hat\rho_M ~ \ra
$$
\be
\label{63}
\ra ~ \hat\rho_{BM}
 ~ \ra ~ \hat\rho_B\bigotimes\hat\rho_M ~ \ra ~
\hat\rho_{BM} ~ \ra ~ \hat\rho_{AM} ~\ra ~ \hat\rho_A \; ,
\ee
where the notation of time, for brevity, is omitted. The convolution
of channels is also a channel.

The set of channels, describing the sequence of time evolutions transforming
the system from the statistical ensemble $\{{\cal H}_B, {\rho}_B\}$ to the
ensemble $\{{\cal H}_A, {\rho}_A\}$, is known
\cite{Brukner_65,Fritz_66,Markovitch_67,Markovitch_68,Emary_69} to be
isomorphic to the statistical ensemble $\{{\cal H}_{AB}, {\rho}_{AB}\}$
of a composite system. The channel-state duality is the Choi-Jamiolkowski
isomorphism \cite{Choi_70,Jamiolkowski_71,Choi_72} which, for the considered
case, yields
\be
\label{64}
 \cH_{AB} \equiv \cH_A \bigotimes \cH_B \;  .
\ee
The equalities
$$
\hat\rho_A \equiv {\rm Tr}_B\hat\rho_{AB} \; , \qquad
\hat\rho_B \equiv {\rm Tr}_A\hat\rho_{AB}
$$
and the normalization conditions
\be
\label{65}
  {\rm Tr}_{AB}\hat\rho_{AB} = {\rm Tr}_{A}\hat\rho_{A} =
{\rm Tr}_{B}\hat\rho_{B} = 1
\ee
are assumed.

Thus, the Choi-Jamiolkowski isomorphism translates the multi-channel picture
of sequentially measuring the observables ${\hat B}$ and ${\hat A}$ into the
consideration of the composite system in the statistical state $\hat\rho_{AB}$.

\section{Probability of composite events}

Since the eigenfunctions of the measured observables ${\hat B}$ and ${\hat A}$
are respectively $|\alpha \rangle$ and $|n \rangle$, the basis in space (64)
is composed of the vectors
\be
\label{66}
 | n\al \rgl \equiv | n \rgl \bigotimes |\al \rgl \;  .
\ee
Hence this space can be represented as
\be
\label{67}
 \cH_{AB} = {\rm span} \{ | n\al \rgl \} \;  .
\ee

Measuring the eigenvalues $B_\alpha$ and $A_n$ corresponds to observing the composite
event $A_n \bigotimes B_\alpha$ represented by the tensor product of two events. The
general mathematical properties of tensor products in measure theory have been studied
in a number of works (see, e.g., \cite{Wilce_73,Foulis_74,Harding_75}).

In our case, the composite event $A_n \bigotimes B_\alpha$ defines the correspondence
\be
\label{68}
  A_n \bigotimes B_\al ~ \ra ~ | n\al \rgl ~ \ra ~
\hat P_n \bigotimes \hat P_\al \; ,
\ee
with the composite projector
\be
\label{69}
 \hat P_n \bigotimes \hat P_\al =   | n\al \rgl \lgl n\al |
\ee
satisfying the resolution
\be
\label{70}
  \sum_{n\al} \hat P_n \bigotimes \hat P_\al = \hat 1_{AB} \; ,
\ee
where ${\hat 1}_{AB}$ is the identity operator in space (67).

The probability of the composite event $A_n \bigotimes B_\alpha$ is defined by
the formula
\be
\label{71}
 p\left (A_n \bigotimes B_\al \right ) \equiv {\rm Tr}_{AB} \hat\rho_{AB}
\hat P_n \bigotimes \hat P_\al \;  ,
\ee
which is a straightforward generalization of definition (10) and which results in
\be
\label{72}
 p \left (A_n \bigotimes B_\al \right ) =
\lgl n\al \; |\;  \hat\rho_{AB}\; |\; n\al \rgl \;  .
\ee
Generally, this probability is not symmetric with respect to the interchange
of $A_n$ and $B_\alpha$, since ${\hat \rho}_{AB}$ may be not the same as
${\hat \rho}_{BA}$.

The probabilities of separate events are given by the marginal forms
\be
\label{73}
p(A_n) = {\rm Tr}_{AB} \hat\rho_{AB}  \hat P_n =
{\rm Tr}_{A} \hat\rho_{A}  \hat P_n \; ,   \qquad
p(B_\al) = {\rm Tr}_{AB} \hat\rho_{AB}  \hat P_\al =
{\rm Tr}_{B} \hat\rho_{B}  \hat P_\al \; ,
\ee
that can also be represented as
$$
 p(A_n) = \sum_\al p\left (A_n\bigotimes B_\al \right) =
\lgl n \;|\; \hat\rho_A \;|\; n \rgl \; ,
$$
\be
\label{74}
p(B_\al) = \sum_n p \left (A_n\bigotimes B_\al \right ) =
\lgl \al\; |\; \hat\rho_B \; |\; \al \rgl \;  .
\ee
In view of resolution (70), the normalization condition holds,
\be
\label{75}
 \sum_{n\al}  p\left ( A_n\bigotimes B_\al \right ) = 1\; .
\ee

The above properties demonstrate that the probability of the composite event
$A_n \bigotimes B_\alpha$, defined by Eq. (71), can be treated as the quantum
joint probability of two events, being valid for arbitrary events, whether
compatible or not. Respectively, for the joint probability, there corresponds
the conditional probability
\be
\label{76}
 p(A_n|B_\al) = \frac{p(A_n\bigotimes B_\al)}{p(B_\al)} \;  ,
\ee
enjoying the standard property of conditional probabilities
\be
\label{77}
 \sum_n p(A_n|B_\al) = 1 \;  .
\ee
Relation (76) is the generalization of the classical Bayes rule for quantum
probabilities, based on the given definition of the quantum joint probabilities.

The most general form of the composite-system state is
\be
\label{78}
  \hat\rho_{AB} =
\sum_{mn} \sum_{\al\bt} \rho_{mn}^{\al\bt} |\; m\al \rgl \lgl n\bt\; | \; ,
\ee
in which
\be
\label{79}
 \left (\rho_{mn}^{\al\bt} \right )^* = \rho_{nm}^{\bt\al} \; ,
\qquad \sum_{n\al} \rho_{nn}^{\al\al} = 1 \; ,
\qquad 0 \leq \rho_{nn}^{\al\al} \leq 1 \;  .
\ee
Then, the probability (72) reads as
\be
\label{80}
 p \left (A_n\bigotimes B_\al \right ) = \rho_{nn}^{\al\al} \;  .
\ee

Let us emphasize that the probability of the factorized event
$A_n \bigotimes B_\alpha$ does not involve interference terms.

\section{Measurements under uncertainty}

Interference terms in decision theory arise when decisions are made under
uncertainty \cite{Yukalov_76,Yukalov_77,Yukalov_78,Yukalov_79,Yukalov_80}.
Similarly, in measurement theory, such terms should arise when there exists
some uncertainty in measurements. Uncertain measurements correspond to
generalized propositions, as has been described above, defined through the
multimode states (41). The final measurement should be operationally testable.
Hence an uncertain measurement can occur only at the intermediate stage of
a measurement procedure. For instance, we can consider the composite prospect
\be
\label{81}
  \pi_n = A_n \bigotimes B \; ,
\ee
consisting of measuring an observable $\hat B$, with not a uniquely defined
result, described by a multimode state (41), and then measuring an observable
$\hat A$, characterized by its operationally testable eigenvalues $A_n$.
The set ${\cal L} \equiv \{\pi_n\}$ forms a prospect lattice. Each composite
prospect (81) is represented by the prospect state
\be
\label{82}
  | \pi_n \rgl = | n \rgl \bigotimes | B \rgl =
\sum_\al b_\al | n \al \rgl \; .
\ee
According to the general prescription, we have the correspondence
\be
\label{83}
  \pi_n ~ \ra ~ |  \pi_n \rgl ~ \ra ~ \hat P( \pi_n) \;  ,
\ee
with the prospect operator
\be
\label{84}
 \hat P( \pi_n) = \hat P_n \bigotimes \hat P_B =
|  \pi_n \rgl \lgl  \pi_n | \;  .
\ee

The prospect states $|\pi_n \rangle$, generally, are not normalized to one
and are not orthogonal to each other. Therefore the prospect operators (84)
are not projectors, since they are not orthogonal to each other and are not
necessarily idempotent,
$$
 \hat P^2( \pi_n) = \lgl  \pi_n |  \pi_n \rgl \hat P(\pi_n) \; .
$$
But the resolution of unity is required, so that
\be
\label{85}
 \sum_n  \hat P( \pi_n)  = \hat 1_{AB} \; .
\ee
By definition (84), the prospect operators are self-adjoint and positive. The family
$\{ {\hat P}(\pi_n) \}$, satisfying condition (85), forms a positive operator-valued
measure.

The prospect probability is
\be
\label{86}
   p( \pi_n) \equiv {\rm Tr}_{AB} \hat\rho_{AB} \hat P(\pi_n) \; ,
\ee
with the properties
\be
\label{87}
 \sum_n  p( \pi_n)  =  1 \; , \qquad 0 \leq p( \pi_n) \leq 1 \;,
\ee
showing that the set $\{ p(\pi_n) \}$ composes a probability measure.

Explicitly, definition (86) gives
\be
\label{88}
p( \pi_n) =
\sum_{\al\bt} b_\al^* b_\bt \lgl n\al \;|\; \hat\rho_{AB}\; |\; n\bt \rgl \;   .
\ee
Separating here the diagonal terms
\be
\label{89}
f(\pi_n) \equiv
\sum_{\al} | b_\al|^2 \lgl n\al\; |\; \hat\rho_{AB}\; |\; n\al \rgl
\ee
from the off-diagonal terms
\be
\label{90}
  q(\pi_n) \equiv
\sum_{\al\neq\bt} b_\al^* b_\bt \lgl n\al\; |\; \hat\rho_{AB}\; |\; n\bt \rgl
\ee
results in the sum
\be
\label{91}
 p(\pi_n) = f(\pi_n) + q(\pi_n) \;  .
\ee
The diagonal and off-diagonal parts can be written as
$$
f(\pi_n) = \sum_{\al} | b_\al|^2 \rho_{nn}^{\al\al} =
\sum_{\al} | b_\al|^2 p \left (A_n \bigotimes B_\al \right ) \; ,
$$
\be
\label{92}
 q(\pi_n) = \sum_{\al\neq\bt} b_\al^* b_\bt \rho_{nn}^{\al\bt} =
2{\rm Re} \sum_{\al<\bt} b_\al^* b_\bt \rho_{nn}^{\al\bt} \; .
\ee

The off-diagonal part (90) or (92) describes the interference due to the occurrence
of the multimode state $|B\rangle$. Such interference effects are typical of quantum
phenomena. The interference term disappears if either the multimode state degenerates
to a single state $|\alpha_0\rgl$ or when the composite-system state is separable,
such that ${\hat \rho}_{AB}$ reads as the diagonal sum
$\sum_{n \alpha} \rho_{nn}^{\alpha \alpha} |n \alpha \rangle \langle n \alpha|$.
Note that the composite-system state is separable only when the measurements of
two observables are not temporally correlated, but correlated measurements
define an entangled composite-system state
\cite{Brukner_65,Fritz_66,Markovitch_67,Markovitch_68,Emary_69}. Thus, the necessary
conditions for the occurrence of interference are the existence in a composite
prospect of uncertainty, corresponding to a multimode state, and the entanglement in
the composite system state, caused by measurement correlations.

Prospect (81) that cannot be reduced to the simple factorized form of two elementary
prospects, but involves the union $B = \bigcup_\alpha B_\alpha$, can be called
{\it entangled prospect}. In the presence of such entangled prospects, the Bayes
rule, introducing the related conditional probability, becomes
$$
 p(A_n|B) \equiv \frac{p(A_n\bigotimes B)}{p(B)} =
\frac{\sum_\al|b_\al|^2p(A_n\bigotimes B_\al)+q(\pi_n)}{\sum_\al|b_\al|^2p(B_\al)+q(B)} \;  .
$$

In agreement with the quantum-classical correspondence principle, the quantum
probability has to reduce to the corresponding classical probability, when quantum
effects, such as interference, disappear. This implies the existence of the limit
\be
\label{93}
 p(\pi_n) ~ \ra ~ f(\pi_n) \; , \qquad  q(\pi_n) ~ \ra ~ 0 \;  ,
\ee
where $f(\pi_n)$ is a classical probability satisfying the standard conditions
\be
\label{94}
 \sum_n f(\pi_n) = 1 \; , \qquad 0 \leq f(\pi_n) \leq 1 \;  .
\ee
Assuming the validity of these conditions in sum (91) requires that the interference
term enjoys the properties
\be
\label{95}
 \sum_n q(\pi_n) = 0 \; , \qquad -1 \leq q(\pi_n) \leq 1 \;   .
\ee

As a simple example of the situation corresponding to a composite event, we may
recall the double-slit experiment. A particle is emitted in the direction of a screen
having two slits. From another side of the screen, there are detectors registering the
arrival of the particle. Let the registration of a particle by a detector number $n$
be denoted as $A_n$ and the passage of the particle through one of the slits be denoted
by $B$, with $B_1$ or $B_2$ being the passage of the particle through the corresponding
slits. When the passage of the particle through a slit $B_\alpha$ is certain, then the
event $A_n \bigotimes B_\alpha$ is factorized and displays no interference, with the
event probability given by $p(A_n \bigotimes B_\alpha)$. However, when it is not known
through which of the slits the particle passes, then the events $\pi_n=A_n\bigotimes B$
are entangled and demonstrate interference. The two-mode state
$|B\rgl = b_1 |\alpha_1 \rangle + b_2 |\alpha_2 \rangle$ is an example of the
multimode states considered above.

\section{Interference in quantum games}

Interference can also arise in the examples related to quantum game theory, where the 
process of measurements under uncertainty is replaced by decision making under uncertainty.    
A typical instance of game theory is provided by the prisoner dilemma game, which
possesses a structure that many other games can be reduced to
\cite{Dresher_82,Rapoport_83,Poundstone_84,Wiebull_85,Kaminski_86}. Here, we consider
the quantum variant of the game in the frame of the above approach.

The generic structure of the prisoner dilemma game is as follows. Two participants can
either cooperate with each other or defect from cooperation. Let the cooperation action
of one of them be denoted by $C_1$ and the defection by $D_1$. Similarly, the cooperation
of the second subject is denoted by $C_2$ and the defection by $D_2$. Depending on their
actions, the participants receive payoffs from the set $\{x_1,x_2,x_3,x_4\}$.

There are four admissible cases: both participants cooperate ($C_1 \bigotimes C_2$),
one cooperates and another defects ($C_1 \bigotimes D_2$), the first defects but the
second cooperates ($D_1 \bigotimes C_2$), and both defect ($D_1 \bigotimes D_2$). The
payoffs to each of them, depending on their actions, are given according to the rule
\begin{eqnarray}
\label{111}
\left [ \begin{array}{ll}
C_1 \bigotimes C_2 &~~ C_1 \bigotimes D_2 \\
D_1 \bigotimes C_2 &~~ D_1 \bigotimes D_2
\end{array} \right ]
~ \longrightarrow ~
\left [ \begin{array}{ll}
x_1  &~~ x_2  \\
x_3  &~~ x_4
\end{array}  \right ]
~ + ~
\left [ \begin{array}{ll}
x_1  &~~ x_3  \\
x_2  &~~ x_4
\end{array}  \right ] \; ,
\end{eqnarray}
where the first (respectively, second) matrix in the r.h.s corresponds to the payoff
of the first (respectively, second) player.

The most interesting question in the game is what choice the participants make, when
they do not know the choice of the other side. This corresponds to the situation where
each of the participants chooses between two prospects. For the first player, these
two prospects are
\be
\label{112}
 \pi_1 =  C_1 \bigotimes \left ( C_2 \bigcup D_2 \right ) \; , \qquad
\pi_2 =  D_1 \bigotimes \left ( C_2 \bigcup D_2 \right ) \; ,
\ee
and, similarly, for the second. Since the game is symmetric with respect to the
players, it is sufficient to consider only one of the players, say the first one. That
is, we need to consider the binary prospect lattice ${\cal L} = \{\pi_1, \pi_2\}$. As
is seen, this situation is the same as in the measurements under uncertainty.
Following the general prescription, we have the following quantum probabilities for
the prospects:
$$
p(\pi_n) = f(\pi_n) + q(\pi_n) \qquad (n=1,2) \; ,
$$
\be
\label{113}
f(\pi_1) = p \left ( C_1 \bigotimes C_2 \right )  +
p \left ( C_1 \bigotimes D_2 \right )\; , \qquad
f(\pi_2) = p \left ( D_1 \bigotimes C_2 \right )  +
p \left ( D_1 \bigotimes D_2 \right )\; .
\ee

The interference factor $q(\pi_n)$, generally, is random, being different for different
participants. If one assumes that the game is realized for a large number of participant
pairs and that all participants have no preference to whether cooperate or defect, when
they are not aware of the action of the other side, then the aggregate interference
factor for each prospect, averaged over all participants, is expected to be zero. This
fact can be called the {\it interference symmetry}.

It is possible to estimate the typical values of the positive and negative interference
factors in the case of a non-informative prior. Assume that these factors are randomly
distributed with a distribution function $\mu(q)$. In view of Eqs. (95), the properties
\be
\label{114}
\int_{-1}^1 \mu(q) \; dq = 1 \; , \qquad
\int_{-1}^1 q\mu(q) \; dq = 0 \; ,
\ee
are valid. Let us denote
\be
\label{115}
 q_+ \equiv  \int_0^1 q \mu(q) \; dq \; , \qquad
q_- \equiv \int_{-1}^0 q\mu(q) \;  .
\ee
Non-informative prior corresponds to the uniform distribution $\mu(q) = 1/2$, which
gives
\be
\label{116}
 q_+ = \frac{1}{4} \; , \qquad q_- = - \; \frac{1}{4} \;  .
\ee
This means that, when the interference symmetry is present, the interference factors
for different players, but for the same prospect, are randomly distributed around
$\pm 0.25$, so that on average their sum is zero. Then the average quantum probability
should coincide with the classical probability $f(\pi_n)$.

If the payoffs are defined so that
\be
\label{117}
 x_3 > x_1 > x_4 > x_2 \;  ,
\ee
then, according to classical utility theory \cite{Neumann_87}, the strategy of
defecting for each player is always more profitable for each of the decision of the
other player, and both players have to defect
\cite{Rapoport_83,Poundstone_84,Wiebull_85,Kaminski_86}. This implies that $f(\pi_1)$
has to be close to zero, while $f(\pi_2)$, close to one.

It is a surprising fact that empirical data, collected for many prisoner dilemma game
realizations \cite{Camerer_88,Tversky_89,Tversky_90}, show that the fraction of those
who choose to cooperate, under the uncertainty of having no information on the choice
of their counterpart, is essentially larger than that prescribed by the classical theory.
Thus, Tversky and Shafir \cite{Tversky_89,Tversky_90} give the empirical fractions
of those who cooperate or defect under uncertainty as $p_{exp}(\pi_1) = 0.37$ and
$p_{exp}(\pi_2) = 0.63$, respectively, which is essentially different from the case
corresponding to the decision under certain information, $f(\pi_1) = 0.1$ and
$f(\pi_2) = 0.9$. The probabilities here are defined as the corresponding fractions
of the participants.

If we apply the rules of the quantum game to humans, then the above data, for the game
under uncertainty, correspond to the effect of {\it spontaneous breaking of interference
symmetry}, so that the interference factor is positive for cooperation and negative for
defection. Then the related quantum probabilities are estimated by the formulas
\be
\label{118}
 p(\pi_1) = f(\pi_1) + 0.25 \; , \qquad
 p(\pi_2) = f(\pi_2) - 0.25 \;  .
\ee
The breaking of symmetry, in the case of humans, is easily understood as the
inclination to cooperation, which is supported by numerous empirical data
\cite{Rapoport_83,Poundstone_84,Wiebull_85,Kaminski_86}, probably
as a result of hard-wired emotional decision modulii that have evolved
over our long evolutionary past as hunter-gatherers cooperating in small groups
\cite{Richerson_91}.

Applying the above formulas to the experiment of Tversky and Shafir
\cite{Tversky_89,Tversky_90}, we find $p(\pi_1) = 0.35$ and $p(\pi_2) = 0.65$.
This, within the accuracy of the experiment, coincides with the results of
Tversky and Shafir, giving $p_{exp}(\pi_1) = 0.37$ and
$p_{exp}(\pi_2) = 0.63$.

We may conclude that the rules of quantum games can be applied to real-life
situations, taking into account that the interference symmetry is broken by
human biases and feelings. For the case of quantum measurements, this would be
analogous to saying that the measurements are not absolutely random, but influence
different prospects in an asymmetric way. For instance, the measuring device can
have a defect that systematically shifts the measured results in one direction.

\section{Multimode quantum systems}

An important application of the developed approach is to defining the quantum 
probabilities of composite events for multi-mode systems. There exist numerous 
realizations of multi-mode quantum systems. These could be atoms with several 
populated electron levels, molecules with several roto-vibrational modes,
quantum dots with several exciton modes, spin systems with several spin projections,
Bose-condensed trapped gases with several coherent modes, and so on \cite{Birman_88}.

Let us consider a multimode quantum system described by a state $|\psi \rangle$
satisfying the Schr\"{o}dinger equation
\be
\label{96}
 i \frac{d}{dt} \; |\psi\rgl = H | \psi\rgl \;  ,
\ee
with a Hamiltonian $H = H(t) = H_0 + V(t)$ composed of a part $H_0$ independent
of time $t$ and a part $V(t)$ depending on $t$. The modes $|n\rangle$ are the
stationary solutions, which are the eigenvectors of the Hamiltonian $H_0$. The
state $|\psi \rangle$ can be expanded over the basis composed of the modes,
\be
\label{97}
 | \psi\rgl = \sum_n c_n | n \rgl  \; .
\ee
Substituting this expansion into Eq. (96) yields the equations for the functions
$c_n = c_n(t)$. The state is normalized, such that
\be
\label{98}
 \lgl \psi | \psi\rgl = \sum_n | c_n |^2 = 1 \;  .
\ee

Suppose we are interested in observing different modes that can be employed for
applications to information processing, quantum chemistry and so on
\cite{Nielsen_1,Keyl_2,Galindo_3,Lupo_53,Shi_53,Tannor_53,Dobek_53}. Assume that,
at the moment of time $t$, we are studying the modes $|n\rangle$ and let us
denote the observation of a mode $|n\rangle$ at time $t$ as an event $A_n$.
And let us denote the observation of a mode $|\alpha\rangle$ at a preceding time
$t_0 < t$ as an event $B_\alpha$. The set of modes at different times could be
different, resulting in different spaces (51). But even if the modes are the same,
it is always admissible to define the spaces (51) as copies of a Hilbert space.

For the composite system in space (67), the matrix elements of the statistical
operator can be written in the form
\be
\label{99}
\rho_{mn}^{\al\bt} \equiv \lgl m \al\; |\; \hat\rho_{AB}\; |\; n \bt \rgl =
c_{m\al} c^*_{n\bt} \;   .
\ee
The probability that at time $t_0$ there happened the event $B_\alpha$, while
at a later time $t > t_0$, the event $A_n$ is observed, according to Eq. (80),
reads as
\be
\label{100}
 p \left ( A_n \bigotimes B_\al \right ) = | c_{n\al}|^2 \;  .
\ee
In order that the relations
$$
p(A_n) = \sum_\al p(A_n \otimes B_\al) \; , \qquad
   p(B_\al) = \sum_n p(A_n \otimes B_\al) \; ,
$$
\be
\label{101}
\sum_n p(A_n) = \sum_\al p(B_\al) = 1
\ee
be valid, the coefficients $c_{n\alpha}$ have to satisfy the normalizations
$$
\sum_\al | c_{n\al} |^2 = | c_n|^2 \; , \qquad
\sum_n | c_{n\al}|^2 = |c_\al|^2 \; ,
$$
\be
\label{102}
\sum_{n\al} |c_{n\al} |^2 = \sum_n | c_n|^2 = \sum_\al |c_\al|^2 = 1 \; ,
\ee
which is in agreement with properties (79).

Now assume that at time $t_0$, an uncertain event $B = \bigcup_\alpha B_\alpha$
happened, with the modes being uniformly weighted, so that $|b_\alpha|^2 = const$,
which can be set to equal one. Looking for the probabilities of the prospects
\be
\label{103}
  \pi_n = A_n \bigotimes \bigcup_\al B_\al \; ,
\ee
we follow the previous section defining the measurements under uncertainty. Then
the diagonal part (89) of the prospect probability gives
\be
\label{104}
 f(\pi_n) =\sum_\al p\left ( A_n \bigotimes B_\al \right ) = |c_n|^2 \;  ,
\ee
while for the interference term (90), we get
\be
\label{105}
q(\pi_n) = \sum_{\al\neq \bt}  b_\al^* b_\bt c_{n\al} c_{n\bt}^* \;  .
\ee

As is mentioned in the previous section, the necessary condition for a nonzero
interference term is the existence of entanglement for state (78). However, this
is not a sufficient condition. State (78) can be entangled, and also generating
entanglement, but the interference term be zero.

For measuring entanglement production, we need to compare the total state (78)
with the reduced operators
$$
\hat\rho_A \equiv {\rm Tr}_B \hat\rho_{AB}  =
\sum_{mn} \sum_\al \rho_{mn}^{\al\al} |m\rgl \lgl n | \;  ,
\qquad
\hat\rho_B \equiv {\rm Tr}_A \hat\rho_{AB}  =
\sum_{n} \sum_{\al\bt} \rho_{nn}^{\al\bt} | \al \rgl \lgl \bt | \;  .
$$
The measure of entanglement production is defined \cite{Yukalov_81} as
\be
\label{106}
 \ep (\hat\rho_{AB}) = \log \;
\frac{||\hat\rho_{AB}||}{||\hat\rho_{A}||\;||\hat\rho_{B}||} ,
\ee
where the spectral norms are assumed, over the basis of the vectors $|n \alpha\rangle$
forming space (67), and, respectively, over the bases of the vectors $|n\rangle$ and
$|\alpha\rangle$. This yields
$$
 ||\hat\rho_{AB}|| =\sup_{n\al} \rho_{nn}^{\al\al} = \sup_{n\al} | c_{n\al} |^2 \;  ,
$$
$$
 ||\hat\rho_{A}|| =\sup_{n}  \sum_\al \rho_{nn}^{\al\al} = \sup_{n} | c_n|^2 \;  ,
\qquad
||\hat\rho_{B}|| =\sup_{\al}  \sum_n \rho_{nn}^{\al\al} = \sup_{\al} | c_\al|^2 \;
$$
Then Eq. (\ref{106}) gives
\be
\label{107}
 \ep (\hat\rho_{AB}) = \log \;
\frac{ \sup_{n\al}|c_{n\al}|^2}{(\sup_n|c_n|^2)(\sup_\al|c_\al|^2) } \;  .
\ee

Let the number of modes be
$$
 M = {\rm dim}\cH_A = {\rm dim}\cH_B \;  .
$$
And let us consider the maximally entangled generalized Bell states, for which
\be
\label{108}
 c_{n\al} = \frac{\dlt_{n\al}}{\sqrt{M} } \;   .
\ee
In this case, the statistical operator of the composite system reduces to
\be
\label{109}
 \hat\rho_{AB} = \frac{1}{M} \; \sum_{mn} | \; mm \rgl \lgl nn \; | \;  .
\ee
In the particular case of two modes, $M = 2$, this corresponds to the standard
Bell state.

For such generalized Bell states, the statistical operator (78) is evidently entangled.
It is also generating entanglement, quantified by measure (\ref{107}) giving
\be
\label{110}
\ep (\hat\rho_{AB}) = \log M \;   .
\ee
However, as is easy to see, the interference term (90) is zero, $q(\pi_n) = 0$.

Thus, the entanglement of the composite state (78) is a necessary, but not sufficient
condition for the occurrence of interference in the probability of a composite
entangled prospect (\ref{103}).

\section{Conclusion}

We have studied the problem of defining quantum probabilities of several events,
so that the quantum probability could be considered as an extension of the
corresponding classical probability. The main results of the article are as follows.

(i) The L\"{u}ders probability of consecutive measurements is a transition
probability between two quantum states and it cannot be accepted as a quantum
extension generalizing the classical conditional probability.

(ii) The Wigner distribution is a weighted L\"{u}ders probability, that is, the
weighted transition probability and it cannot be treated as a quantum extension of
the classical joint probability.

(iii) Quantum joint probabilities can be introduced as probabilities of composite
events represented by tensor products of events that can be of any nature, whether
compatible or incompatible.

(iv) The probability of measurements under uncertainty is defined by employing
positive operator-valued measures.

(v) The necessary condition for the appearance of an interference term in a
quantum probability is that the considered prospect be entangled and the system
state be entangled. This condition is necessary, but not sufficient.

(vi) Applying the approach to quantum games demonstrates the occurrence of the
spontaneous interference symmetry breaking.

(vii) The approach is used for characterizing multi-mode systems. Such systems are 
ubiquitous in a variety of physical applications.   

The developed approach can be employed in the theory of quantum measurements and
quantum decision theory that is a part of the measurement theory. It can be useful 
for creating artificial quantum intelligence \cite{Yukalov_100}. Among important
physical applications is the probabilistic description of multi-mode systems. Here
we have considered the general scheme for a multi-mode system of arbitrary nature.
The concrete example of a trapped Bose-condensed atomic system with several coherent 
modes will be presented in a separate paper.    

\vskip 2mm

{\bf Acknowledgment}

Financial support from the Swiss National Science Foundation is appreciated.
We are very grateful to J.R. Busemeyer and E.P. Yukalova for many useful
discussions.


\newpage


\begin{thebibliography}{99}

\bibitem{Nielsen_1}
Nielsen M and Chuang I 2000
{\it Quantum Computation and Quantum Information}
(Cambridge: Cambridge University)

\bibitem{Keyl_2}
Keyl M 2002
{\it Phys. Rep.} {\bf 369} 431

\bibitem{Galindo_3}
Galindo A and Martin-Delgado A 2002
{\it Rev. Mod. Phys.} {\bf 74} 347

\bibitem{Born_4}
Born M 1926
{\it Zeit. Physik} {\bf 37} 863

\bibitem{Reed_5}
Reed M and Simon B 1972
{\it Methods of Mathematical Physics} Vol. 1
(New York: Academic)

\bibitem{Neumann_6}
von Neumann J 1955
{\it Mathematical Foundations of Quantum Mechanics}
(Princeton: Princeton University)

\bibitem{Luders_7}
L\"{u}ders G 1951
{\it Ann. Physik} {\bf 15} 663

\bibitem{Wigner_8}
Wigner E 1932
{\it Phys. Rev.} {\bf 40} 749

\bibitem{Nelson_9}
Nelson E 1967
{\it Dynamical Theories of Brownian Motion}
(Princeton: Princeton University)

\bibitem{Gudder_10}
Gudder S 1979
{\it Stochastic Methods in Quantum Mechanics}
(New York: North-Holland)

\bibitem{Fine_11}
Fine A 1982
{\it Phys. Rev. Lett.} {\bf 48} 291

\bibitem{Hall_12}
Hall M J 1989
{\it Found. Phys.} {\bf 19} 189

\bibitem{Gudder_13}
Gudder S 1990
{\it Found. Phys.} {\bf 20} 499

\bibitem{Malley_14}
Malley J D 2004
{\it Phys. Rev. A} {\bf 69} 022118

\bibitem{Malley_15}
Malley J D and Fine A 2005
{\it Phys. Lett. A} {\bf 347} 51

\bibitem{Malley_16}
Malley J D 2006
{\it Phys. Lett. A} {\bf 359} 122

\bibitem{Niestegge_17}
Niestegge G 2004
{\it J. Math. Phys.} {\bf 45} 4714

\bibitem{Niestegge_18}
Niestegge G 2008
{\it Found. Phys.} {\bf 38} 241

\bibitem{Bohr_19}
Bohr N 1913
{\it Philos. Mag.} {\bf 26} 1, 476, 857

\bibitem{Wheeler_20}
Wheeler J A and Zurek W H 1983
{\it Quantum Theory and Measurement}
(Princeton: Princeton University)

\bibitem{Zurek_21}
Zurek W H 2003
{\it Rev. Mod. Phys.} {\bf 75} 715

\bibitem{Birkhoff_22}
Birkhoff G and von Neumann J 1936
{\it Ann. Math.} {\bf 37} 823

\bibitem{Benioff_23}
Benioff P A 1972
{\it J. Math. Phys.} {\bf 13} 908

\bibitem{Holevo_24}
Holevo A S 1973
{\it J. Math. Anal.} {\bf 3} 337

\bibitem{Yukalov_25}
Yukalov V I 2002
{\it Phys. Rev. E} {\bf 65} 056118

\bibitem{Yukalov_26}
Yukalov V I 2003
{\it Phys. Lett. A} {\bf 308} 313

\bibitem{Yukalov_27}
Yukalov V I 2012
{\it Phys. Lett. A} {\bf 376} 550

\bibitem{Bogolubov_28}
Bogolubov N N 1967
{\it Lectures on Quantum Statistics} Vol. 1
(New York: Gordon and Breach)

\bibitem{Bogolubov_29}
Bogolubov N N 1970
{\it Lectures on Quantum Statistics} Vol. 2
(New York: Gordon and Breach)

\bibitem{Gleason_30}
Gleason A M 1957
{\it J. Math. Mech.} {\bf  6} 885

\bibitem{Kirkwood_31}
Kirkwood J G 1933
{\it Phys. Rev.} {\bf 44} 31

\bibitem{Caves_32}
Caves C M, Fuchs C A and Schack R 2002
{\it Phys. Rev. A} {\bf 65} 022305

\bibitem{Leifer_33}
Leifer M S 2006
{\it Phys. Rev. A} {\bf 74} 042310

\bibitem{Leifer_34}
Leifer M S and Poulin D 2008
{\it Ann. Phys. (N.Y.)} {\bf 323} 1899

\bibitem{Fuchs_35}
Fuchs C A and Schack R 2011
{\it Found. Phys.} {\bf 41} 345

\bibitem{Wallace_35}
Wallace D 2013
arXiv:1306.4907

\bibitem{Margenau_36}
Margenau H 1963
{\it Ann. Phys. (N.Y.)} {\bf 23}, 469 (1963).

\bibitem{Moladuer_37}
Moldauer P A 1972
{\it Found. Phys.} {\bf 2} 41

\bibitem{Ludwig_38}
Ludwig G 1983
{\it Foundations of Quantum Mechanics} (Berlin: Springer)

\bibitem{Braginsky_39}
Braginsky V B and Khalili F Y 1996
{\it Rev. Mod. Phys.} {\bf 68} 1

\bibitem{Merkli_40}
Merkli M, Sigal I M and Berman G P 2008
{\it Ann. Phys. (N.Y.)} {\bf 323} 373

\bibitem{Merkli_41}
Merkli M, Berman G P and Sigal I M 2008
{\it Ann. Phys. (N.Y.)} {\bf 323} 3091

\bibitem{Yukalov_42}
Yukalov V I 2012
{\it Ann. Phys. (N.Y.)} {\bf 327} 253

\bibitem{Kadison_43}
Kadison R V 1951
{\it Ann. Math.} {\bf 54} 325

\bibitem{Polkovnikov_44}
Polkovnikov A 2010
{\it Ann. Phys. (N.Y.)} {\bf 325} 1790

\bibitem{Bell_45}
Bell J S 1990
{\it Phys. World} {\bf 3} 33

\bibitem{Barchielli_45}
Barchielli A, Lanz L and Prosperi G M 1982
{\it Nuovo Cimento B} {\bf 72} 79

\bibitem{Ghirardi_45}
Ghirardi G C, Rimini A and Weber T 1986
{\it Phys. Rev. D} {\bf 34} 470

\bibitem{Pearle_45}
Pearle P 1989
{\it Phys. Rev. A} {\bf 39} 2277

\bibitem{Johansen_46}
Johansen L M 2007
{\it Phys. Rev. A} {\bf 76} 012119

\bibitem{Busemeyer_47}
Busemeyer J R and Bruza P 2012
{\it Quantum Models of Cognition and Decision}
(Cambridge: Cambridge University)

\bibitem{Randall_48}
Randall C H and Foulis D J 1983
{\it Found. Phys.} {\bf 13} 843

\bibitem{Yuen_49}
Yuen H P 1982
{\it Phys. Lett. A} {\bf 91} 101

\bibitem{Huttner_50}
Huttner B, Muller A, Gauter J D, Zbinden H and Gisin N 1996
{\it Phys. Rev. A} {\bf 54} 3783

\bibitem{Suppes_51}
Suppes P 1966
{\it Philos. Sci.} {\bf 33} 14

\bibitem{Putnam_52}
Putnam H 1969
{\it Boston Stud. Philos. Sci.} {\bf 5} 199

\bibitem{Lupo_53}
Lupo D W and Quack M 1987
{\it Chem. Rev.} {\bf 87} 181

\bibitem{Shi_53}
Shi S, Woody A and Rabitz H 1988
{\it J. Chem. Phys.} {\bf 88} 6870

\bibitem{Tannor_53}
Tannor D J and Rice S A 1988
{\it Adv. Chem. Phys.} {\bf 70} 441

\bibitem{Dobek_53}
Dobek K, Karpinski M, Demkowicz-Dobrzanski R, Banaszek K and Horodecki P 2013
{\it Laser Phys.} {\bf 23} 025204

\bibitem{Davies_54}
Davies E B 1976
{\it Quantum Theory of Open Systems} (New York: Academic)

\bibitem{Kraus_55}
Kraus K 1983
{\it States, Effects, and Operations} (Berlin: Springer)

\bibitem{Holevo_56}
Holevo A S 2001
{\it Statistical Structure of Quantum Theory} (Berlin: Springer)

\bibitem{Foulis_57}
Foulis D, Piron C and Randal C 1983
{\it Found. Phys.} {\bf 13} 813

\bibitem{Foulis_58}
Foulis D J, Greechie R J and R\"{u}ttimann G T 1992
{\it Int. J. Theor. Phys.} {\bf 31} 789

\bibitem{Yukalov_59}
Yukalov V I 1970
{\it Moscow Univ. Phys. Bull.} {\bf 25} 49

\bibitem{Yukalov_60}
Yukalov V I 1971
{\it Moscow Univ. Phys. Bull.} {\bf 26} 22

\bibitem{Gelfand_61}
Gelfand I M and Neumark M A 1943
{\it Mat. Sbornik} {\bf 12} 197

\bibitem{Kraus_62}
Kraus K 1971
{\it Ann. Phys. (N.Y.)} {\bf 64} 311

\bibitem{Holevo_63}
Holevo A S 2011
{\it Probabilistic and Statistical Aspects of Quantum Theory}
(Berlin: Springer)

\bibitem{Holevo_64}
Holevo A S and Giovannetti V 2012
{\it Rep. Prog. Phys.} {\bf 75} 046001

\bibitem{Brukner_65}
Brukner C, Taylor S, Cheung S and Vedral V 2004
arXiv:quant-ph/0402127

\bibitem{Fritz_66}
Fritz T 2010
{\it New J. Phys.} {\bf 12} 083055

\bibitem{Markovitch_67}
Markovitch S and Reznik B 2011
arXiv:1103.2557

\bibitem{Markovitch_68}
Markovitch S and Reznik B 2011
arXiv:1107.2186

\bibitem{Emary_69}
Emary C, Lambert N and Nori F 2013
arXiv:1304.5133

\bibitem{Choi_70}
Choi M D 1972
{\it Can. J. Math.} {\bf 24} 520

\bibitem{Jamiolkowski_71}
Jamiolkowski A 1972
{\it Rep. Math. Phys.} {\bf 3} 275

\bibitem{Choi_72}
Choi M D 1975
{\it Lin. Alg. Appl.} {\bf 10} 285

\bibitem{Wilce_73}
Wilce A 1992
{\it Int. J. Theor. Phys.} {\bf 31} 1915

\bibitem{Foulis_74}
Foulis D J and Bennett M K 1993
{\it Order} {\bf 10} 271

\bibitem{Harding_75}
Harding J 2009
{\it Int. J. Theor. Phys.} {\bf 48} 769

\bibitem{Yukalov_76}
Yukalov V I and Sornette D 2008
{\it Phys. Lett. A} {\bf 372} 6867

\bibitem{Yukalov_77}
Yukalov V I and Sornette D 2009
{\it Eur. Phys. J. B} {\bf 71} 533

\bibitem{Yukalov_78}
Yukalov V I and Sornette D 2009
{\it Entropy} {\bf 11} 1073

\bibitem{Yukalov_79}
Yukalov V I and Sornette D 2010
{\it Adv. Complex Syst.} {\bf 13} 659

\bibitem{Yukalov_80}
Yukalov V I and Sornette D 2011
{\it Theor. Decis.} {\bf 70} 283

\bibitem{Dresher_82}
Dresher M 1961
{\it The Mathematics of Games of Strategy: Theory and Applications}
(Englewood Cliffs: Prentice-Hall)

\bibitem{Rapoport_83}
Rapoport A and Chammah A M 1965
{\it Prisoner's Dilemma} (Ann Arbor: University of Michigan)

\bibitem{Poundstone_84}
Poundstone W 1992
{\it Prisoner's Dilemma} (New York: Doubleday)

\bibitem{Wiebull_85}
Weibull J W 1995
{\it Evolutionary Game Theory} (Cambridge: Massachusetts Institute of Technology)

\bibitem{Kaminski_86}
Kaminski M 2004
{\it Games Prisoners Play} (Princeton: Princeton University)

\bibitem{Neumann_87}
von Neumann J and Morgenstern O 1953
{\it Theory of Games and Economic Behavior} (Princeton: Princeton University)

\bibitem{Camerer_88}
Camerer C 2003
{\it Behavioral Game Theory} (Princeton: Princeton University)

\bibitem{Tversky_89}
Tversky A and Shafir E 1992
{\it Psychol. Sci.} {\bf 3} 305

\bibitem{Tversky_90}
Tversky A 2004
{\it Preference, Belief, and Similarity: Selected Writings}
(Cambridge: Massachusetts Institute of Technology)

\bibitem{Richerson_91}
Richerson P J and Boyd R 2006
{\it Not by Genes Alone: How Culture Transformed Human Evolution}
(Chicago: University of Chicago)

\bibitem{Birman_88}
Birman J L, Nazmitdinov R G and Yukalov V I 2013
{\it Phys. Rep.} {\bf 526} 1

\bibitem{Yukalov_81}
Yukalov V I 2003
{\it Phys. Rev. A} {\bf 68} 022109

\bibitem{Yukalov_100}
Yukalov V I and Sornette D 2009
{\it Laser Phys. Lett.} {\bf 6} 833


\end{thebibliography}
\end{document}